\documentclass[aps, reprint, prl, twocolumn]{revtex4-1}
\usepackage{graphicx}
\usepackage{amsmath}
\usepackage{bm}
\usepackage{braket}
\usepackage{psfrag} 
\usepackage{latexsym} 
\usepackage{amstext}
\usepackage{amsxtra} 
\usepackage{float}
\usepackage{upgreek} 
\usepackage{times} 
\usepackage[hidelinks,bookmarks=false]{hyperref}
\usepackage{pgffor}
\usepackage{pdfpages}

\allowdisplaybreaks

\makeatletter
\AtBeginDocument{\let\LS@rot\@undefined}
\makeatother

\begin{document}

\newcommand{\kB}{k_{\rm B}}
\newcommand{\cs}{$\clubsuit \; $}
\newcommand{\downstate}{\left\vert\downarrow\right\rangle}  
\newcommand{\upstate}{\left\vert\uparrow\right\rangle}
\newcommand{\overbar}[1]{\mkern 1.5mu\overline{\mkern-1.5mu#1\mkern-1.5mu}\mkern 1.5mu}
\newcommand{\expect}[1]{\langle#1\rangle}
\newcommand{\lb}{\ell_B}
\newcommand{\vh}{v_\textrm{d}}
\newcommand{\vhVec}{\vec{v}_\textrm{d}}
\newcommand{\affcua}{MIT-Harvard Center for Ultracold Atoms, Research Laboratory of Electronics, and Department of Physics, Massachusetts Institute of Technology, Cambridge, Massachusetts 02139, USA}
\newcommand{\Vs}{V_\textrm{s}}
\newcommand{\sLLL}{\sigma_\textrm{LLL}}
\newcommand{\nll}{N_\textrm{LL}}

\title{
Geometric squeezing into the lowest Landau level
}

\author{Richard~J.~Fletcher}
\author{Airlia~Shaffer}
\author{Cedric~C.~Wilson}
\author{Parth~B.~Patel}
\author{Zhenjie~Yan}
\author{Valentin Cr\'{e}pel}
\author{Biswaroop~Mukherjee}
\author{Martin~W.~Zwierlein}
\affiliation{\affcua}

\begin{abstract}
The equivalence between neutral particles under rotation and charged particles in a magnetic field relates phenomena as diverse as spinning atomic nuclei, weather patterns, and the quantum Hall effect.
In their quantum descriptions, translations along different directions do not commute, implying a Heisenberg uncertainty relation between spatial coordinates. 
Here, we exploit the ability to squeeze non-commuting variables to dynamically create a Bose-Einstein condensate occupying a single Landau gauge wavefunction in the lowest Landau level.
We directly resolve the extent of the zero-point cyclotron orbits, and demonstrate geometric squeezing of the orbits' guiding centers by more than ${7}~$dB below the standard quantum limit.
The condensate attains an angular momentum of more than ${1000}\,{\hbar}$ per particle, and an interatomic distance comparable to the size of the cyclotron orbits. This offers a new route towards strongly correlated fluids and bosonic quantum Hall states.
\end{abstract}

\date{\today}



\maketitle

In 1851, Foucault directly demonstrated the rotation of the Earth via the precession of a pendulum's oscillation axis. 
This occurs because in the rotating frame, counter- and co-rotating motions no longer oscillate at the pendulum's natural frequency, $\omega$. Instead, their frequencies are increased and decreased respectively by the Earth's rotation frequency, $\Omega$, which leads to the bob performing epicycles as illustrated in Fig.~1A.
In Foucault's experiment, for which $\Omega\ll\omega$, this manifests as an apparent precession of the oscillation axis.
If we imagine instead that $\Omega=\omega$, the centrifugal force exactly cancels the restoring force. 
While the pendulum can still perform cyclotron orbits against the frame's rotation, the motion of the orbit's guiding center is free. 
In a quantum-mechanical description, the energy spectrum is closely analogous to that of charged particles in a magnetic field. 
It forms discrete Landau levels spaced by $2\hbar\omega$, corresponding to different states of cyclotron motion, each with a large degeneracy arising from the possible guiding center positions.

\begin{figure*}[t] 
   \centering
   \includegraphics[width=2\columnwidth]{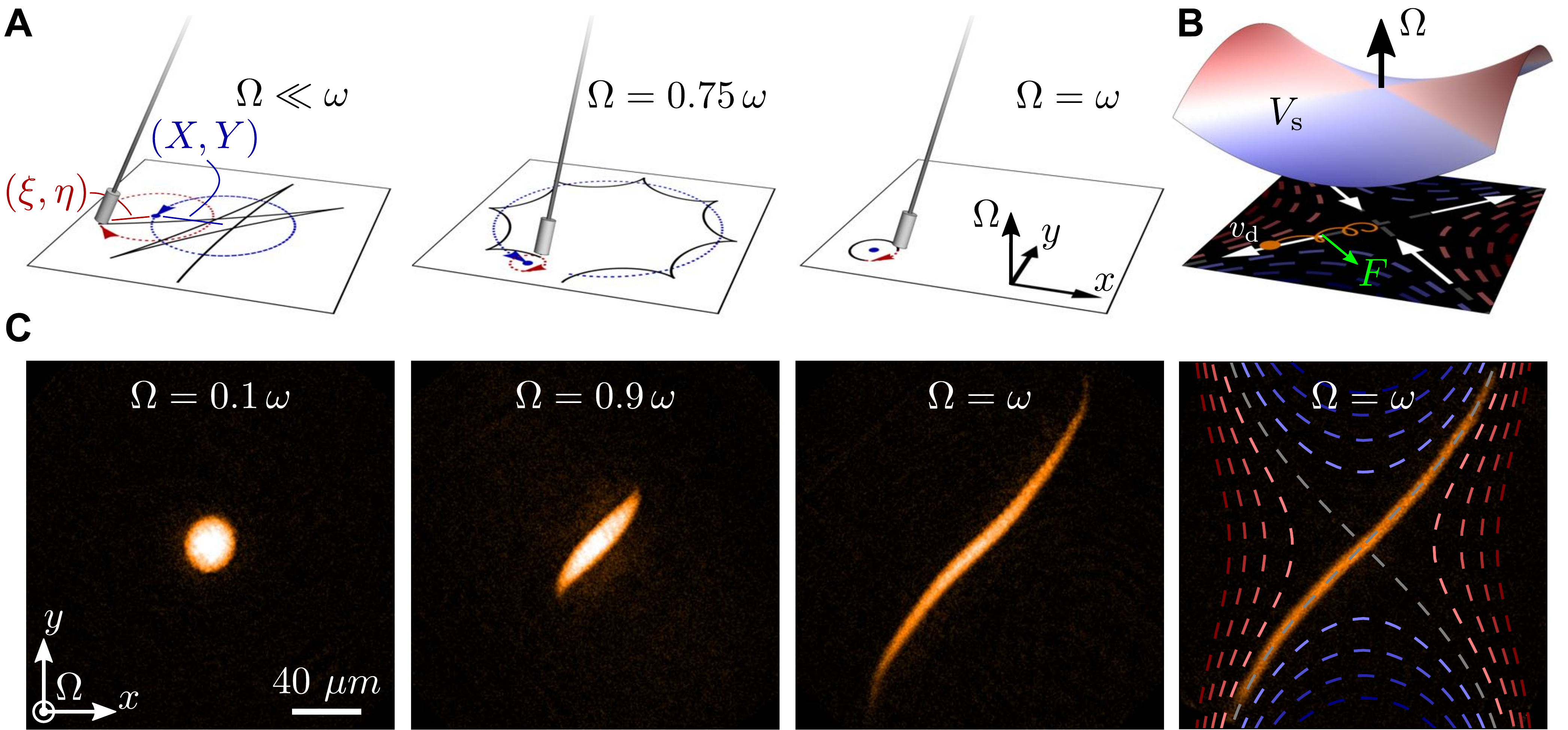} 
   \caption{
   \textbf{Geometric squeezing of a rotating Bose-Einstein condensate.}
   \textbf{(A)}~Viewed in a frame rotating at $\Omega$, the motion of a Foucault pendulum with natural frequency $\omega$ separates into a slow co-rotating drift of the guiding center $(X,Y)$, shown in blue, and fast counter-rotating cyclotron orbits with relative coordinates $(\xi,\eta)$, shown in red. For $\Omega<\omega$ the pendulum performs skipping orbits, while if $\Omega=\omega$ the guiding center motion is free. 
    \textbf{(B)}~Atoms in an elliptical harmonic trap rotating at $\Omega=\omega$ evolve under both a vector potential and a scalar saddle potential $V_\textrm{s}$, whose isopotentials are shown by red ($V_\textrm{s}>0$) and blue ($V_\textrm{s}<0$) dashed lines. Particles perform cyclotron orbits, whose guiding centers drift along isopotentials with a velocity $\vhVec$ (white arrows) orthogonal to the local force $\vec{F}=-\vec{\nabla} V_\textrm{s}$ (green arrow).
   \textbf{(C)}~\emph{In situ} images of the condensate in the rotating frame. During the hold time at $\Omega=\omega$, the atoms flow out along one diagonal and in along the other, mediating squeezing of the distribution in guiding center phase-space. The final image is overlaid with the isopotentials of $\Vs$.}
\end{figure*}

An intrinsic characteristic of both neutral particles under rotation and charged particles in a magnetic field is the non-commutativity of space. This can be seen from the quantized Hamiltonian of a pendulum of mass $m$ viewed in the rotating frame,
\begin{align}
\hat{H}&=\frac{\hat{p}_x^2+\hat{p}_y^2}{2 m}+\frac{1}{2}m\omega^2(\hat{x}^2+\hat{y}^2)-\Omega \hat{L}_z,
\label{eqn:rotatingHO}
\end{align}
where $\hat{p}_{x,y}$ are the canonical momenta along $x$ and $y$, and $\hat{L}_z$ is the axial angular momentum. The rotational term $\Omega \hat{L}_z$ mixes spatial and momentum coordinates into new normal modes, and one decouples Eq.~(\ref{eqn:rotatingHO}) by transforming into cyclotron coordinates 
$\xi=\frac{1}{2}\left(x-\frac{p_y}{m\omega}\right)$ and 
$\eta=\frac{1}{2}\left(y+\frac{p_x}{m\omega}\right)$, and guiding center coordinates 
$X=\frac{1}{2}\left(x+\frac{p_y}{m\omega}\right)$ and
$Y=\frac{1}{2}\left(y-\frac{p_x}{m\omega}\right)$, yielding~\cite{SI}
\begin{align}
\hat{H}&=m\omega(\omega+\Omega)(\hat{\xi}^2+\hat{\eta}^2)+m\omega(\omega-\Omega)(\hat{X}^2+\hat{Y}^2).
\label{eq:H}
\end{align}
Since $\hat{x}=\hat{X}+\hat{\xi}$ and $\hat{y}=\hat{Y}+\hat{\eta}$, the particle's motion is the sum of a fast cyclotron motion and a slow drift of the guiding center (see Fig.~1A).
Crucially, while the absolute spatial coordinates $\hat{x}$ and $\hat{y}$ always commute, the two pairs of cyclotron and guiding center coordinates separately do not. Each pair spans the phase space of a one-dimensional harmonic oscillator, and consequently
\begin{equation}
[\hat{\xi},\hat{\eta}]=-[\hat{X},\hat{Y}]=i\lb^2,
\end{equation}
where $\lb=\sqrt{\hbar/(2m\omega)}$ is the rotational analogue of the magnetic length. 
If an applied potential $\hat{V}(\hat{x},\hat{y})$ varies little over this lengthscale, it cannot resolve the cyclotron motion and only couples to the guiding centers. In this case $\hat{V}(\hat{x},\hat{y})\rightarrow \hat{V}(\hat{X},\hat{Y})$, and the resulting dynamics occurs within a non-commutative space~\cite{Wilkinson:1987}.

This non-commutativity of guiding center motion lies at the heart of the Hall effect. Each spatial variable generates translations in the orthogonal direction, meaning that a force along $X$ effects motion along $Y$.
Particles therefore drift along isopotentials of $V$ with a velocity $\vec{v}_\textrm{d}=\vec{\Omega}\times\vec{\nabla} V/(2m\Omega\omega)$ in analogy to the $\vec{E}\times\vec{B}$ drift of electromagnetism. This flow is divergence-free, reflecting the incompressibility of phase-space distributions~\cite{Note1},
and defines a one-to-one mapping between a particle's initial and final position. Time evolution therefore always results in a purely geometric, equiareal transformation of the guiding center distribution.

These concepts are relevant to atomic nuclei~\cite{Valatin:1956,Bohr:1976,Mottelson:1976}, astrophysical phenomena~\cite{kerr:1963,Maeder:2012}, quantum Hall systems~\cite{Haldane:2011}, and ultracold atomic quantum gases, which offer a highly versatile experimental arena for studying rotating quantum fluids~\cite{Cooper:2008}.
In Bose-Einstein condensates rotating close to the trap frequency, signatures of the gas approaching the lowest Landau level (LLL) were seen in a softening of the vortex lattice~\cite{Schweikhard:2004a,Bretin:2004}.
A principal goal is to address the quantum Hall regime, but the exacting requirements on the trap isotropy and rotation speed present a major challenge.
Synthetic magnetic fields~\cite{dalibard:2011,goldman:2014, galitski:2019} have also been engineered by other methods, such as spin-orbit coupling~\cite{Lin:2009c,Galitski:2013}, and by direct phase-imprinting in both optical lattices~\cite{struck:2012,aidelsburger:2013, miyake:2013,jotzu:2014} and synthetic dimensions~\cite{celi:2014}. 
Experiments demonstrated a transverse Hall response in both lattice transport~\cite{aidelsburger:2014} and superfluid collective modes~\cite{leblanc:2012}, and chiral edge states in synthetic dimensions~\cite{stuhl:2015, mancini:2015}.

Here, we directly exploit the non-commutativity of guiding center motion to realize geometric squeezing, cleanly distilling a single Landau gauge wavefunction in the lowest Landau level~\cite{SI}.
In comparison to previous work in azimuthally-symmetric condensates~\cite{Schweikhard:2004a}, this obviates delicate fine-tuning of trapping and rotation parameters, and offers a complementary `Landau gauge' starting point from which to investigate interaction-driven physics in quantum Hall systems.
To begin our experiment, we prepare a condensate of $N_\textrm{Tot}=8.1(1)\times10^5$ atoms of $^{23}$Na in an elliptical time-orbiting-potential (TOP) trap~\cite{Petrich:1995}, with trap frequencies $(\omega_x,\omega_y,\omega_z)=(\sqrt{1+\varepsilon},\sqrt{1-\varepsilon},\sqrt{8})\,\omega$. Here $\omega=2\pi\times88.6(1)~$Hz and the trap ellipticity is $\varepsilon=0.125(4)$. 
We smoothly ramp the trap's rotation frequency from zero to $\omega$, wait for a variable time $t$, and then obtain an absorption image of the \emph{in situ} density distribution. 
Our imaging resolution is sufficient to observe vortices \emph{in situ} with a contrast of $\sim 60\%$~\cite{SI}. These have a characteristic size set by the healing length, which is $\sim 300~$nm in our system. This is significantly smaller than the quantum-mechanical ground state size of cyclotron orbits, set by the magnetic length $\lb=1.6~\mu$m.

In the frame rotating at $\Omega$, the condensate evolves under two distinct potentials.
First, the frame rotation induces a vector potential and hence a synthetic magnetic field~\cite{SI}.
Second, the centrifugal force and TOP trap give rise to a scalar potential $V=m(\omega^2-\Omega^2)(X^2+Y^2)/2+m\varepsilon \omega^2(X^2-Y^2)/2$. 
For $\Omega/\omega<\sqrt{1-\varepsilon}$ the isopotentials of $V$ are closed; the condensate remains confined, but deforms into an ellipse. In earlier experiments, unstable density modulations mediated the nucleation of vortices for rotation frequencies $\Omega/\omega\gtrsim 0.8$~\cite{Madison:2001,Sinha:2001}. By ramping sufficiently quickly we preclude breakup of the condensate, while allowing its ellipticity to adiabatically follow the equilibrium value~\cite{Recati:2001}. 

When $\Omega\:{=}\:\omega$, the scalar potential forms a saddle $\Vs=m\varepsilon \omega^2(X^2-Y^2)/2$, illustrated in Fig.~1B. Without a vector potential, atoms would be lost along the anti-trapped $y$-direction. Instead, the guiding centers drift outward along the $x=y$ contours, and inward along the $x=-y$ contours. This flow is illustrated by white arrows, and mediates squeezing of the spatial distribution.
In Fig.~1C we show the evolution of the condensate density viewed in the rotating frame. The final image is overlaid with the known isopotentials of $\Vs$, whose coincidence with the atomic density provides a qualitative signature of isopotential drift. The small curvature of the diagonal contours arises from the known quartic corrections $\sim(X^2+Y^2)^2$ to the trapping potential~\cite{Petrich:1995}, and the spatial twisting of the condensate lies in close analogy to the twisting in optical phase-space induced by the Kerr effect~\cite{Kirchmair:2013}.

\begin{figure}[t] 
   \centering
   \includegraphics[width=\columnwidth]{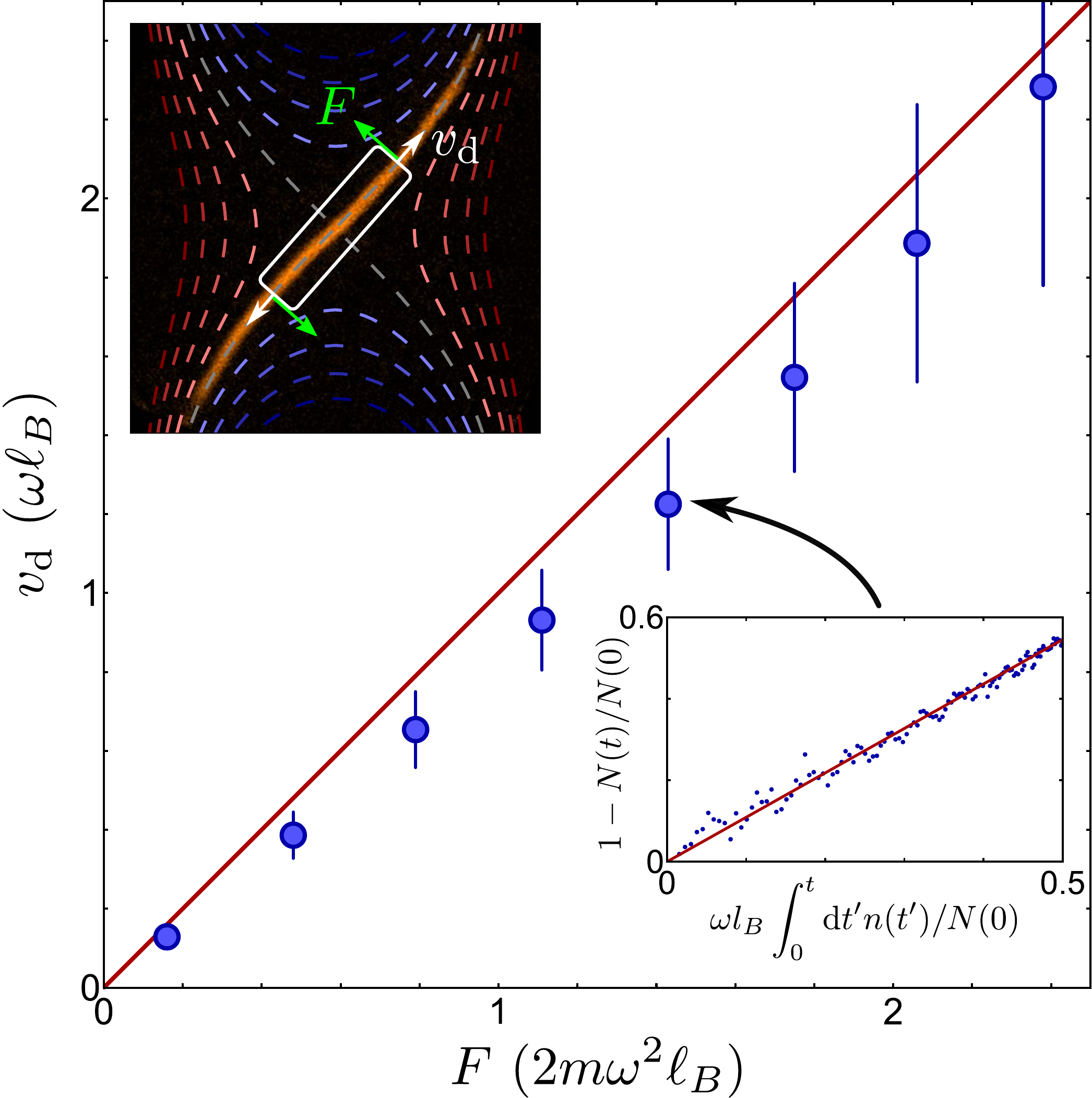} 
   \caption{
   \textbf{Isopotential drift velocity.}
  The main plot shows the radial speed of particles, $\vh$, in response to the azimuthal force, $F$. The speed is inferred from changes in the atom number, $N$, inside a bounding box (top inset), and the density, $n$, at its boundary. The bottom inset shows a typical plot constructed from $N(t)$ and $n(t)$, whose slope gives $\vh$ (see text).
    The data show good agreement with the theoretical expectation (red line) without any free parameters. While the force $F$ is calculated assuming a harmonic trap, quartic corrections to the potential reduce the velocity along $x=y$ giving a small downward shift of the data, which is captured by a GP simulation~\cite{SI}. Error bars show the variation in $\vh$ measured across different time intervals. 
   }
   \label{fig:2}
\end{figure}

To measure the transverse Hall response, we obtain the radial drift speed as a function of the azimuthal force, which at a radius $r$ is $F(r)=m\varepsilon \omega^2 r$. Our measurements are shown in Fig.~2 along with the theoretical relation $\vh=F/(2m\omega)$, valid for any quantum state, which shows good agreement without any free parameters. We infer the drift speed using a continuity equation; the atom number $N$ inside a box (see inset) centered on $r=0$ and with length $2R$ varies as $\dot{N}=-2\vh n$, where $\vh$ and $n$ are the drift speed and one-dimensional number density at $r=R$. Integrating once gives $1-N(t)/N(0)=2\vh \int_0^t\textrm{d}t'n(t')/N(0)$, allowing straightforward evaluation of $\vh$ as shown in the lower inset. This method offers a convenient protocol for measuring the Hall response of any fluid.

While the drift velocity determines the local response to a force, the specific geometric transformation of the cloud depends upon the global shape of $\Vs$. Qualitatively, isopotential flow on a saddle in the presence of a magnetic field results in elongation and contraction along orthogonal diagonals. More quantitatively, in terms of the oscillator ladder operators $\hat{a}=\sqrt{m\omega/\hbar}\;(\hat{\xi}+i\hat{\eta})$ and $\hat{b}=\sqrt{m\omega/\hbar}\;(\hat{X}-i\hat{Y})$ the single-particle Hamiltonian is~\cite{SI}
\begin{equation}
\hat{H}_\textrm{s}\approx
2\hbar\omega\Big(\hat{a}^\dagger\hat{a}+1/2\Big)
+
\frac{\hbar\zeta}{2}\Big(\hat{b}\hat{b}+\hat{b}^\dagger\hat{b}^\dagger\Big),
\end{equation}
where we define $\zeta=\varepsilon \omega/2$.
Comparison with the one-mode squeezing operator $\hat{S}(\alpha)=\exp([\alpha^\ast \hat{b}\hat{b}-\alpha\hat{b}^\dagger\hat{b}^\dagger]/2)$ reveals that time-evolution under a saddle potential is equivalent to fully coherent squeezing of the guiding center phase-space distribution, analogous to phase-squeezing in quantum optics~\cite{Loudon:1987,Fertig:1987,Vishveshwara:2010}. Consistent with the perspective based on isopotential flow, the imaginary squeezing parameter $\alpha=i\zeta t$ describes dilation of the cloud along the diagonals of phase space by factors $\exp(\pm\zeta t)$.

\begin{figure*}[t] 
   \centering
   \includegraphics[width=2\columnwidth]{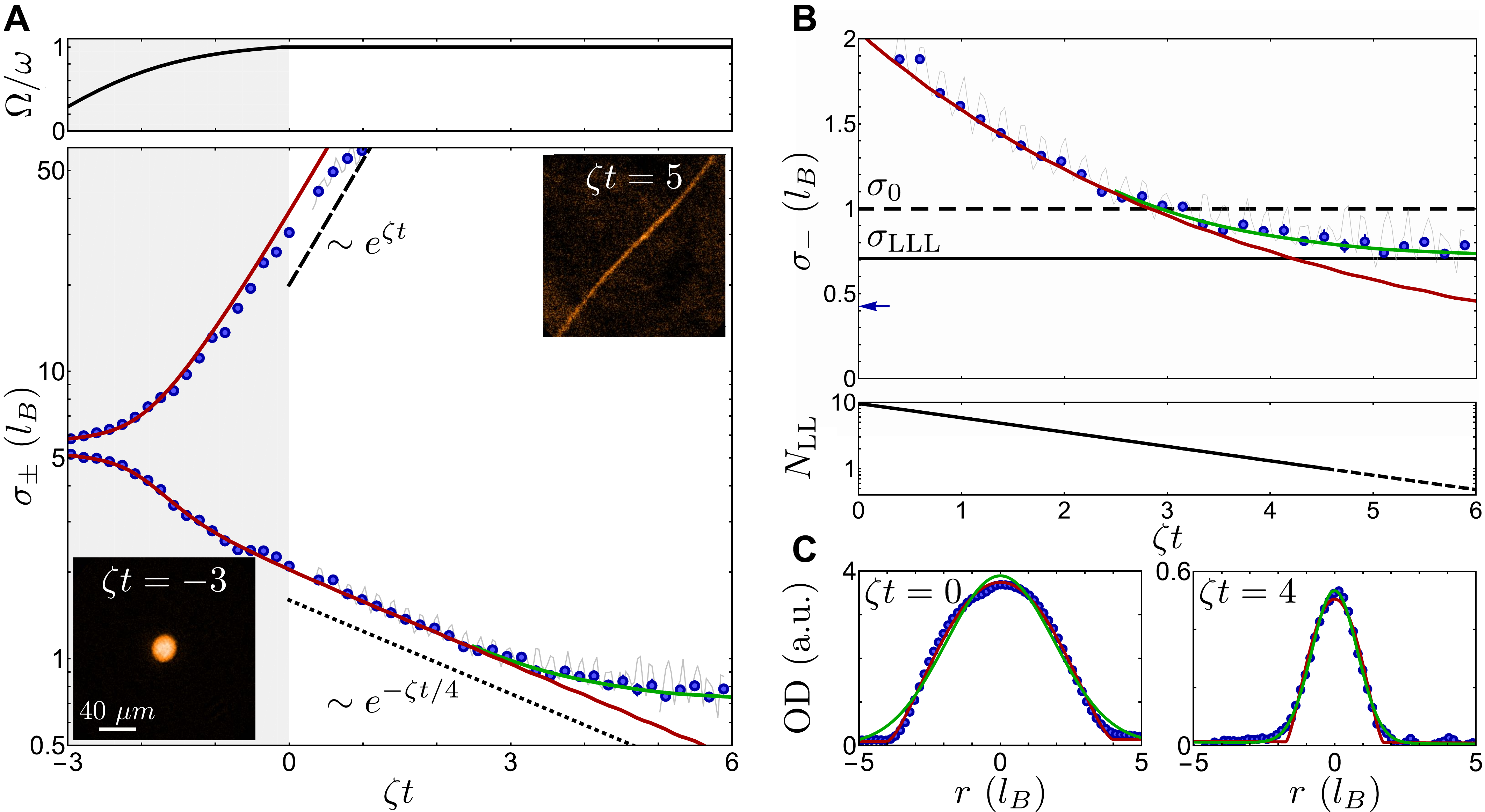} 
   \caption{
   \textbf{Squeezing into the lowest Landau level.}
   \textbf{(A)} Evolution of the major and minor cloud radii, $\sigma_\pm$, with insets showing representative \emph{in situ} images of the cloud. Initially the condensate is approximately isotropic, while for long times the spatial aspect ratio exceeds $100$.  The red line shows the prediction of a hydrodynamic model for which the atom number is the only free parameter, and whose behavior when $\Omega=\omega$ follows simple scalings shown by the dashed and dotted lines (see text). The green line shows the result of a Gross-Pitaevskii simulation of our experiment~\cite{SI}, and captures the deviation from classical hydrodynamic behavior as the LLL is approached. The gray data show a small cyclotron breathing oscillation driven by trap imperfections (see text), while the blue points are averaged over one period.
   \textbf{(B)} The inferred number of occupied Landau levels $N_\textrm{LL}\equiv\mu/(2\hbar\omega)$, along with a zoom-in of the minor width evolution. As the condensate enters the LLL, we observe that its width saturates at $\sLLL=\lb/\sqrt{2}$, shown by a solid line and corresponding to zero-point cyclotron motion. For comparison, the dashed line shows the width of the two-dimensional harmonic oscillator ground state, $\sigma_0=\lb$. The blue arrow denotes the measured imaging resolution obtained using vortex cores~\cite{SI}.
   \textbf{(C)} The transverse optical density (OD) profile of the cloud along with fits of Thomas-Fermi (red) and Gaussian (green) functions. At early times, interactions dominate and the profile is Thomas-Fermi in character, whereas when $N_\textrm{LL}\lesssim 1$ we observe a Gaussian shape, which is characteristic of wavefunctions in the LLL.
   }
   \label{fig:3}
\end{figure*}

In the limit $\zeta t\gg 1$, the particles' guiding centers become widely distributed along one diagonal and sharply localized along the other. 
The residual transverse width of the cloud solely arises from the unsqueezed cyclotron orbits, which have a size $\sqrt{\langle \hat{\xi}^2\rangle}=\lb \sqrt{\nu+1/2}$ in the $\nu^\textrm{th}$ Landau level.
The minimum orbit size $\sLLL=\lb/\sqrt{2}$ occurs in the LLL, where the cyclotron wavefunction is Gaussian and saturates the Heisenberg uncertainty relation $\Delta \xi\Delta \eta\geq \lb^2/2$. 
The density of any condensate in the LLL is therefore a convolution of the guiding center distribution with a Gaussian of width $\sLLL$. In the quantum optics analogy, this directly realizes the Husimi-Q representation of the guiding center Wigner function~\cite{SI}.
In our case, at long times the cloud is an extended strip of transverse width $\sLLL$. Geometric squeezing therefore coherently transforms the condensate into a single Landau gauge wavefunction within the LLL~\cite{SI}.

In Fig.~3A we show images of the condensate before and after squeezing, and plot the major and minor cloud widths, $\sigma_{\pm}$, which are defined as the $e^{-1/2}$-radii obtained from a Gaussian fit. 
Initially, the chemical potential is $\mu_0\approx h\times 3.4~$kHz and the number of Landau levels admixed into the condensate wavefunction is $\sim \mu_0/(2\hbar\omega)\approx 20$, hence the evolution is well-described by a hydrodynamic model which neglects quantum pressure~\cite{Sinha:2001}. The prediction of this model is shown by the red line, for which the only free parameter is the atom number~\cite{Note2}. 

For times $t>0$, the cloud evolves under the squeezing Hamiltonian of Eq.~(3) and the major width increases as $\sigma_+\propto \exp(\zeta t)$, illustrated by the dashed line.
However, the minor width decays more slowly. This difference arises because the condensate size contains contributions from both the guiding centers, which are squeezed at a rate $\zeta$, and from the cyclotron orbits, whose size depends upon the number of occupied Landau levels $N_\textrm{LL}\equiv\mu/(2\hbar\omega)$. 
In our experiment, $\sigma_-$ is generally dominated by cyclotron motion and its evolution is captured well by a simple scaling model. 
The chemical potential is proportional to the atomic number density $\sim N_\textrm{Tot}/(\sigma_+\sigma_-\sigma_z)$, where $\sigma_z$ is the axial extent of the condensate. The major width always increases as $\sigma_+\propto \exp(\zeta t)$, and $\sigma_{-,z}\propto\sqrt{\mu}$ when $N_\textrm{LL}\gg1$. We therefore predict a time-dependence $\sigma_-\propto \exp(-\zeta t/4)$ at early times, which is shown by the dotted line in Fig.~3A. 
The gray data show a small breathing of the cloud at the cyclotron frequency $2\omega$. This is driven by imperfections in the trap, which shows a $\sim0.3\%$ rms variation in $\omega$ with ellipse orientation, giving a perturbation in the rotating frame with a frequency $2\Omega$. The blue points are averaged over one period.
 
The falling chemical potential $\mu\propto \exp(-\zeta t/2)$ guarantees that eventually $\mu<2\hbar\omega$ and the condensate enters the LLL. As shown in Fig.~3B, we directly observe the saturation of $\sigma_-$ at the zero-point cyclotron width $\sLLL$ imposed by Heisenberg uncertainty. 
Since the hydrodynamic model neglects quantum pressure, it predicts that $\sigma_-\rightarrow 0$. On the other hand, the saturation of the cloud width is captured very well by a Gross-Pitaevskii simulation with no free parameters (green solid line)~\cite{SI}.
For comparison, the dashed line shows the width $\sigma_0=\lb$ of the non-interacting harmonic oscillator ground state, which corresponds to minimal, but isotropic, Heisenberg uncertainity in both cyclotron and guiding center coordinates. This lies above our data at long times, and from the last five data we infer squeezing of the guiding centers by $>7~\textrm{dB}$ below the standard quantum limit. 

In the second panel of Fig.~3B we plot the number of occupied Landau levels, $N_\textrm{LL}$, inferred from the central density evaluated using the fitted hydrodynamic model. We indeed find that the crossover to LLL behavior occurs for $N_\textrm{LL}\sim 1$; the dashed region corresponds to $N_\textrm{LL}<1$ where the hydrodynamic model is not applicable and this inference is no longer self-consistent.
We also see a qualitative change in the shape of the cloud, which changes from a Thomas-Fermi to a Gaussian profile. This is shown in Fig.~3C, where we plot cuts along $x=-y$ at early and late times. If $N_\textrm{LL}\gg1$, the healing length is much smaller than the magnetic length and the density profile is a Thomas-Fermi function (red line). On the other hand, if $N_\textrm{LL}<1$ the profile is Gaussian (green line) reflecting the cyclotron ground state.
We note that at our latest times, the interparticle distance has grown to about $500~$nm, close to half the size of a zero-point cyclotron orbit $\sim\sLLL$. This signals the approach of the Bose gas towards the strongly correlated regime~\cite{Cooper:2001,Sinha:2005,Cooper:2008,Chen:2012,Senthil:2013,Vishwanath:2013,galitski:2019}.

Microscopically, the squeezing operator mixes higher angular momentum states into the condensate wavefunction, in analogy to the admixing of higher Fock states in squeezed light~\cite{Loudon:1987}. In general, the angular momentum of a superfluid can either be carried by vortices, or by deformations which break rotational symmetry~\cite{Zambelli:2001}. Here, since $\vec{\nabla}\times\vhVec=0$ the induced flow is irrotational, but the large aspect ratio gives a moment of inertia $\Theta=mN_\textrm{Tot}( \sigma_+^2-\sigma_-^2)^2 /( \sigma_+^2+\sigma_-^2) \approx mN_\textrm{Tot} \sigma_+^2$ which is close to the rigid body value~\cite{Zambelli:2001}. For clouds with $\sigma_+ > 50~\lb$ this gives a per-particle angular momentum $\langle l_z\rangle>1000\,\hbar$ despite the absence of any vortices inside the condensate~\cite{aftalion:2009}.

In the experiments above, geometric squeezing was seen in the evolution of the condensate widths. To directly observe the drift velocity field inside the cloud, we now introduce a dilute gas of vortices which correspond to nodes in the atomic wavefunction and can serve as `tracer particles' for the local flow.
\begin{figure}[t] 
   \centering
   \includegraphics[width=\columnwidth]{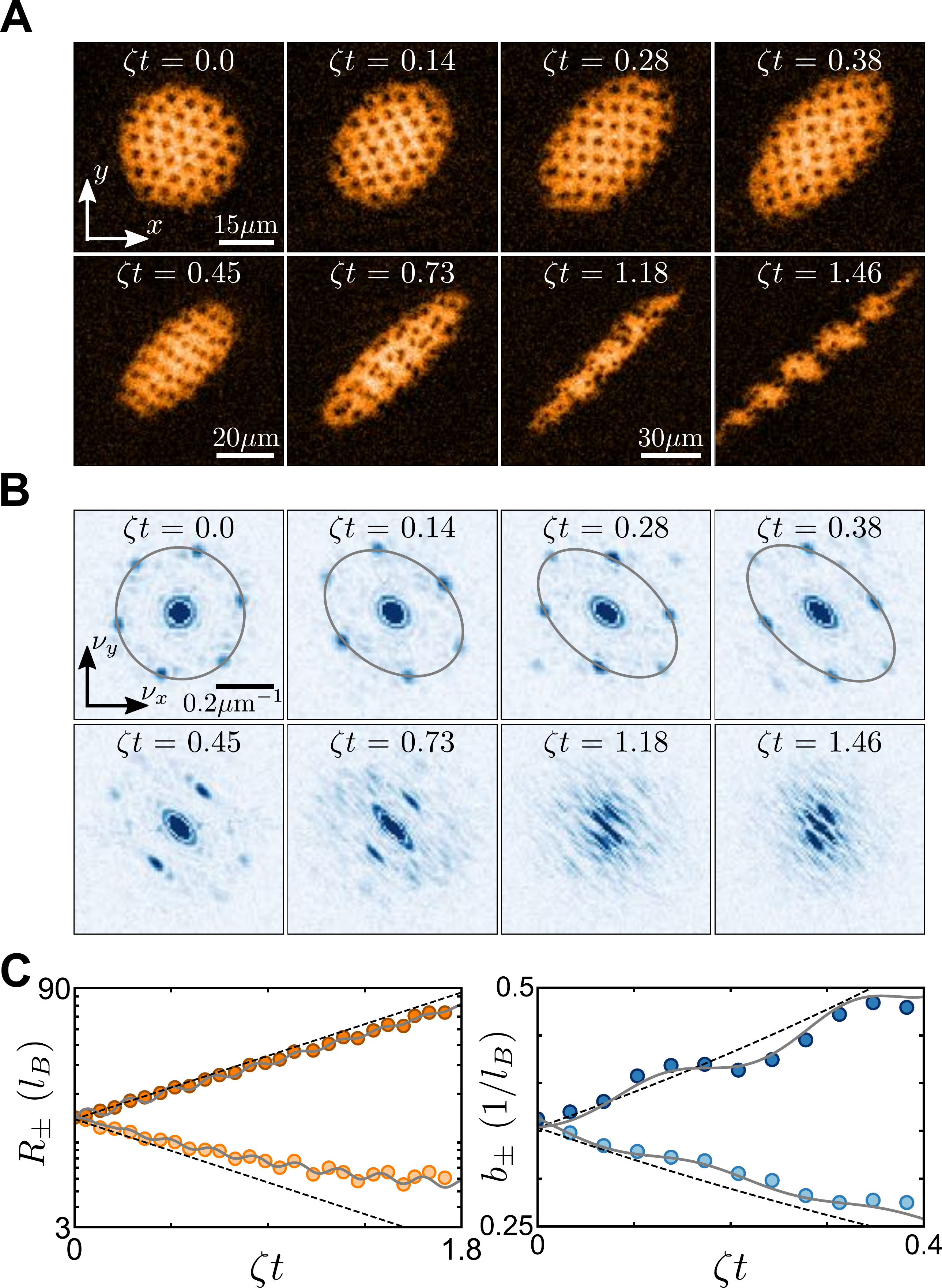} 
   \caption{
   \textbf{Squeezing of a vortex lattice.}
   \textbf{(A-B)}~\emph{In situ} evolution in both real space (top) and reciprocal space (bottom) after suddenly applying the rotating saddle. Initially the cloud is round, and the reciprocal lattice vectors lie on a circle. Squeezing is evident in both the condensate spatial envelope and the vortex lattice spacing. At longer times, clustering of vortices causes the condensate to break up into droplets.
 \textbf{(C)}~Time evolution of the major/minor Thomas-Fermi radii of the condensate, $R_\pm$, and the major/minor radii of the ellipse describing the lattice vectors, $b_\pm$. The black dashed lines show exponential functions $A\exp(\pm\zeta t)$, with $A$ fixed by the data at $t=0$, while the solid lines include the small contributions of quadrupolar collective modes and the non-zero size of the cyclotron orbits~\cite{SI}. 
 The longest squeezing time $\zeta t=1.8$ corresponds to ${t\approx 50~}$ms. The ellipse widths in reciprocal space are shown for times for which the distribution of vortices remains periodic.
   }
   \label{fig:4}
\end{figure}
We prepare a ground state condensate rotating at $0.8\,\omega$ in an isotropic trap, and instantaneously apply the saddle $\Vs$ rotating at $\Omega=\omega$. The initial chemical potential is $\mu\approx h\times 2.2~$kHz giving a cyclotron orbit size $\sim \sqrt{\mu/(2\hbar\omega)}\,\lb= 5.5~\mu$m which is much smaller than the cloud's Thomas-Fermi radius of $21~\mu$m, meaning that the observed width is dominated by the guiding center distribution. In Fig.~4 we show the \emph{in situ} evolution in both real and reciprocal space. Initially, the condensate is circular and contains a triangular Abrikosov lattice with sixfold-symmetric reciprocal lattice vectors. Subsequently, squeezing is evident in both the cloud shape and in the vortex lattice. Since the vortices are distributed throughout the whole cloud, this indicates that the coordinates of all particles evolve under the same squeezing transformation. 
For longer times, while the overall spatial envelope continues to squeeze, the density profile exhibits an intricate evolution. 
Squeezing of the initially triangular vortex lattice eventually leads to the formation of vortex rows~\cite{Engels:2002,Cozzini:2003b}.
Subsequently, a hydrodynamic instability drives amalgamation of the vortices into clusters, and an intriguing fragmentation of the condensate into a persistent array of droplets.

In Fig.~4C, we show the evolution of the major and minor Thomas-Fermi radii of the cloud, $R_{\pm}$, and the major and minor radii of an ellipse fitted to the reciprocal lattice vectors, $b_{\pm}$. The dashed lines show exponential functions $A\exp(\pm\zeta t)$, where the amplitude $A$ is the only free parameter, which capture the initial evolution well. This confirms both the expected rate of squeezing, and the incompressibility of the guiding center distribution. 
The solid lines show a fit which includes the excitation of quadrupolar collective modes by the saddle turn-on and trap imperfections (see above), and additionally accounts for the non-zero cyclotron orbit size~\cite{SI}. This results in a slight reduction of the apparent squeezing rate, and a slowdown of the decay in $R_-$ as the guiding center width approaches the cyclotron size.

The geometric squeezing protocol established here offers a new route to LLL physics in quantum gases. Crucially, simply turning off the saddle potential halts the outward flow of atoms. This controllably prepares an equilibrium condensate~\cite{SI}, which occupies a single Landau gauge wavefunction, whose purely interaction-driven evolution in the flat single-particle dispersion of the LLL can then be cleanly observed.
Natural immediate directions concern the collective excitation spectrum~\cite{Girvin:1985}, quantum hydrodynamic stability, and the appearance of strongly-correlated bosonic states~\cite{Cooper:2001,Oktel:2004,Sinha:2005,aftalion:2009,Chen:2012,Senthil:2013,Vishwanath:2013}. 
More generally, the ability to resolve cyclotron motion and vortices \emph{in situ} allows the study of chiral edge states and quantum turbulence in rotating gases. From a metrology perspective, azimuthally-squeezed condensates might offer benefits for rotation sensing, and a route to spin-squeezing via a spatially-dependent coupling between internal atomic states~\cite{Wineland:1994}.

We thank Tarik Yefsah and Julian Struck for early contributions to the planning and construction of the lab space and apparatus, and Lev Kendrick and J. Thatcher Chamberlin for experimental assistance. 
This work was supported by
the National Science Foundation (Center for Ultracold Atoms
Award No. PHY-1734011, and Award No. PHY- 1506019), 
Air Force Office of Scientific Research (FA9550-16-1-0324 and
MURI Quantum Phases of Matter FA9550-14-1-0035), 
Office of Naval Research (N00014-17-1-2257), 
the DARPA A-PhI program through ARO grant W911NF-19-1-0511, 
the David and Lucile Packard Foundation, 
and the Vannevar Bush Faculty Fellowship.
R.J.F. acknowledges support from the MIT Pappalardo Fellowship.
A.S. acknowledges support from the NSF GRFP.

Correspondence
and requests for materials should be addressed to rfletch@mit.edu.



\section*{Supplementary material}

# Geometric squeezing into the lowest Landau level
# Supplementary Materials

Richard J. Fletcher, Airlia Shaffer, Cedric C. Wilson, Parth B. Patel,
Zhenjie Yan, Valentin Crépel, Biswaroop Mukherjee, and Martin W. Zwierlein

*MIT-Harvard Center for Ultracold Atoms, Research Laboratory of Electronics, and Department
of Physics, Massachusetts Institute of Technology, Cambridge, Massachusetts 02139, USA*

(Dated: March 8, 2021)

## EXPERIMENTAL PROCEDURES

### Sample preparation

We prepare a Bose-Einstein condensate of $N_{\text{Tot}} = 8.1(1) \times 10^5$ atoms of $^{23}$Na in the $|F = 2, m_F = 2\rangle$ hyperfine state, in a magnetic TOP trap [1] of rms radial frequency $\omega = 2\pi \times 88.6(1)$ Hz, and with no discernible thermal component. The chemical potential at the cloud center is $\mu_0 \approx h \times 3.4$ kHz corresponding to a healing length of $\sqrt{\hbar^2/(2m\mu)} = 250$ nm.

### Imaging calibration

The objective used for our *in situ* imaging has a nominal numerical aperture of NA = 0.5. In addition to the diffraction limit, an imaged cloud can also be broadened by optical aberrations, imperfections in the polarisation and frequency of the imaging light, and by motion of the atoms during imaging.

To directly characterize our imaging resolution experimentally, we measure the core structure of a quantum vortex. In Fig. S1A we show a rotating condensate prepared in a circular magnetic trap which has a radial trapping frequency $\omega = 2\pi \times 88.6(1)$ Hz. The cloud rotation rate is $\Omega \approx 0.75\,\omega$, which is determined from the two-dimensional vortex number density, $n_v$, according to $n_v = m\Omega/(\pi\hbar)$ [2]. From the measured Thomas-Fermi radius of the condensate, $R \approx 22.1\,\mu$m, and using the effective radial trapping frequency of $\sqrt{\omega^2 - \Omega^2}$, we infer a central chemical potential $\mu \approx h \times 1.9$ kHz corresponding to a healing length of 340 nm.

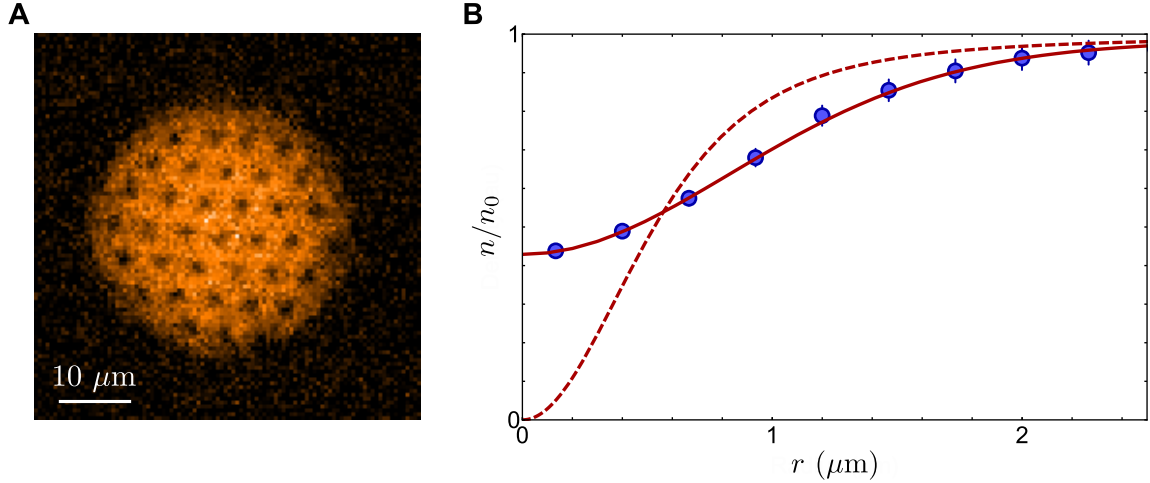

**Fig. S1. *In situ* image of quantum vortices.** **(A)** An *in situ* image of a condensate containing a vortex lattice. A radial average of the measured density, $n$, is performed around the core of each vortex within $R/2$ of the cloud center, where $R$ is the Thomas-Fermi radius of the cloud. **(B)** The average of these profiles is shown by the blue points, where the density, $n$, has been normalized to the density of the surrounding condensate, $n_0$. We fit the data with a function obtained by convolving the theoretical core structure of a single vortex (see text) with a Gaussian function, whose $e^{-1/2}$-radius has an optimal value of 670 nm. The resulting curve is shown by the solid line, while the dashed line shows the theoretical curve without any broadening.

We perform an azimuthal average of the measured two-dimensional atomic number density, $n$, around every vortex located within a radius of $R/2$. The average of these individual vortex profiles is shown in Fig. S1B, where we compare the data to

two models. The dashed curve shows the theoretical vortex core stucture, obtained by numerically solving the Gross-Pitaevskii equation for a single vortex within a uniform condensate [3]. The solid curve shows a fit function obtained by convolving this profile with a Gaussian of variable $e^{-1/2}$-radius $\sigma$, which simultaneously captures both the increased width and reduced contrast of the vortex well. The optimum value of $\sigma = 670$ nm corresponds to the effective broadening of a point source arising from both the diffraction limit and imaging imperfections. It is indicated by a blue arrow in Fig. 3 of the main paper.

For comparison, we measured the point spread function of our imaging system before installation in the machine, using a point source provided by a SNOM optical fibre tip [4]. A Gaussian fitted to this function has a $e^{-1/2}$-radius of 280 nm. The difference between this value and the observed resolution is approximately accounted for by motion of the atoms during imaging; we use a high-intensity imaging pulse with a duration of 3 $\mu$s, resulting in diffusion by $\approx 570$ nm of a $^{23}$Na atom transverse to the imaging axis.

We note that *in situ* detection of vortices has also been reported using, i) dark-field imaging [5], where several vortices within a lattice were detected with a spatial resolution $\sim 3$ $\mu$m and separation of 9 $\mu$m, ii) phase-contrast imaging [6], where individual vortices could be detected for separations $\gtrsim 20$ $\mu$m, and iii) by filling the vortex cores with atoms prepared in a different internal hyperfine state [7], which expanded the core to a radius of $\sim 7$ $\mu$m. In contrast to these previous methods, our imaging allows us to resolve dense vortex lattices, with an inter-vortex spacing comparable to the vortex size, given by the healing length scale of $\sim 0.5$ $\mu$m.

### Fit function for vortex lattice squeezing

In Fig. 4C of the main paper, we show the evolution under geometric squeezing of the major and minor widths of the cloud in real space, and the major and minor widths of an ellipse describing the reciprocal vectors of the vortex lattice. While the dashed lines show purely the ideal squeezing evolution $R_\pm, b_\pm \sim \exp(\pm\zeta t)$, the solid line fits include a small admixture of $m = 0$ and $m = 2$ quadrupolar collective excitations, which have a natural frequency $2\omega$ for a cloud rotating at $\Omega = \omega$ [8]. These are excited by both the sudden switch-on of the saddle potential which initiates squeezing of the cloud, and a small ($\sim 0.3\%$) breathing of the trapping frequency $\omega$ as the trap rotates (see main paper). We also include the non-zero size of cyclotron orbits; these are not squeezed and so their contribution to the cloud width remains constant, while the guiding centers are squeezed at a rate $\zeta$. We simultaneously fit the function

$$R_\pm(t) = A\sqrt{e^{\pm 2\zeta t} + B^2} + C\sin(2\omega t + \phi_C) \pm D\sin(2\omega t + \phi_D)),$$
$$b_\pm(t) = E/R_\pm(t),$$

to all data sets $\{R_\pm(t), b_\pm(t)\}$.

We find fractional amplitudes $C/A = 0.047(5)$ and $D/A = 0.019(5)$ for the $m = 0$ and $m = 2$ modes respectively, and a residual cyclotron orbit size $AB = 7.6(4)$ $\mu$m, comparable to an estimate $\sim \sqrt{\mu/(2\hbar\omega)}\,\ell_B = 5.5$ $\mu$m obtained from the chemical potential $\mu \approx h \times 2.2$ kHz.

### Geometric squeezing as a route to equilibrium rotating gases

Turning off the saddle potential during geometric squeezing halts the outward flow of atoms. The density of the final condensate can be smoothly varied by changing the duration of the squeezing, and is conveniently parameterized by the number of occupied Landau levels $N_{LL} = \mu/(2\hbar\omega)$. The condensate continues to rotate at $\Omega = \omega$, experiencing a synthetic magnetic field but no scalar potential. In the limit $N_{LL} \ll 1$ the atoms occupy a single Landau gauge wavefunction within the LLL, whereas in the limit $N_{LL} \gg 1$ states from the first $\sim N_{LL}$ Landau levels are admixed into the superfluid wavefunction.

In both cases, despite the absence of any scalar potential in the rotating frame, the cloud is stabilized by an effective trap along $x$, of frequency $2\omega$, where we take the long axis of the condensate to lie along the $y$-direction. This can be seen directly from the Hamiltonian in the Landau gauge of a gas rotating at $\Omega = \omega$ (see Eq. (17) below),

$$H = \frac{p_x^2}{2m} + \frac{1}{2}m(2\omega)^2(x - k_y\ell_B^2)^2, \tag{1}$$

where $k_y$ is the wavevector along the translationally-invariant $y$-direction. Physically, the effective trapping arises from the kinetic energy cost imposed by irrotationality of the condensate in the lab frame, which implies a flow profile in the rotating frame $\vec{v} = (0, -2\omega x)$ [9].

In Fig. S2 we show several examples of condensate evolution after turning off the saddle. The cloud continues to rotate at $\Omega = \omega$, but the density profile remains constant. In the rotating frame, the condensate provides a static, equilibrium starting point from which to investigate quantum Hall physics.

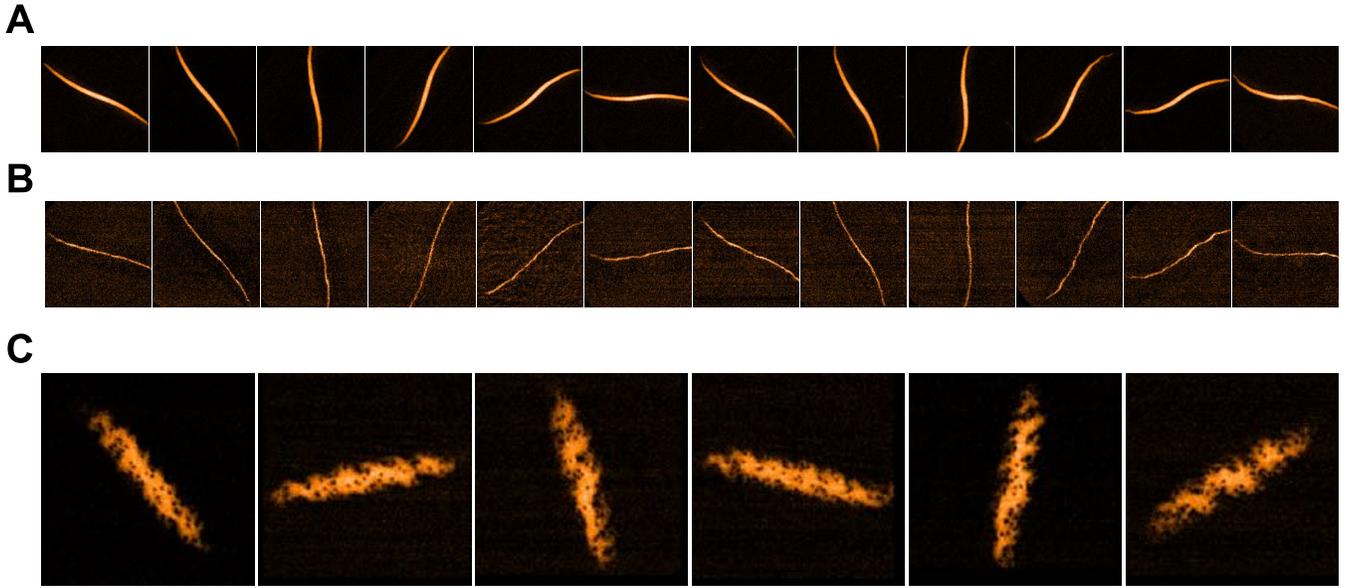

**Fig. S2. Preparation of an equilibrium, rotating condensate by geometric squeezing.** Turning off the saddle potential halts geometric squeezing, and freezes the outward flow of atoms. The cloud continues to rotate at $\Omega = \omega$ and therefore experiences a synthetic magnetic field. Even though it feels no scalar potential, the cloud maintains its shape thanks to the imprinted flow profile in the rotating frame (see text), realizing an equilibrium starting point from which to investigate quantum Hall physics. **(A-B)** The stable evolution of a cloud with $\mu/(2\hbar\omega) \approx 9$, and a lower density cloud with $\mu/(2\hbar\omega) \approx 3$. The time interval between images is 1 ms, with the first image being taken immediately after turning off the saddle potential. **(C)** The evolution of a cloud supporting quantum vortices, obtained by geometric squeezing of a condensate containing a vortex lattice. The images show the cloud 0 ms, 2 ms, 5 ms, 7 ms, 10 ms, and 20 ms after turning off the saddle.

## Gross-Pitaevskii simulations

Given the experimental results presented in the main paper, it is interesting to explore the experimental protocol within a Gross-Pitaevskii (GP) framework. Stimulated by the experiment, we thus implemented a GP simulation in the rotating frame. In addition to describing well the saturation of the cloud width as the gas enters the LLL (see Fig. 3 of the main paper), we also reproduced several other features of the experiment, examples of which are shown in Fig. S3.

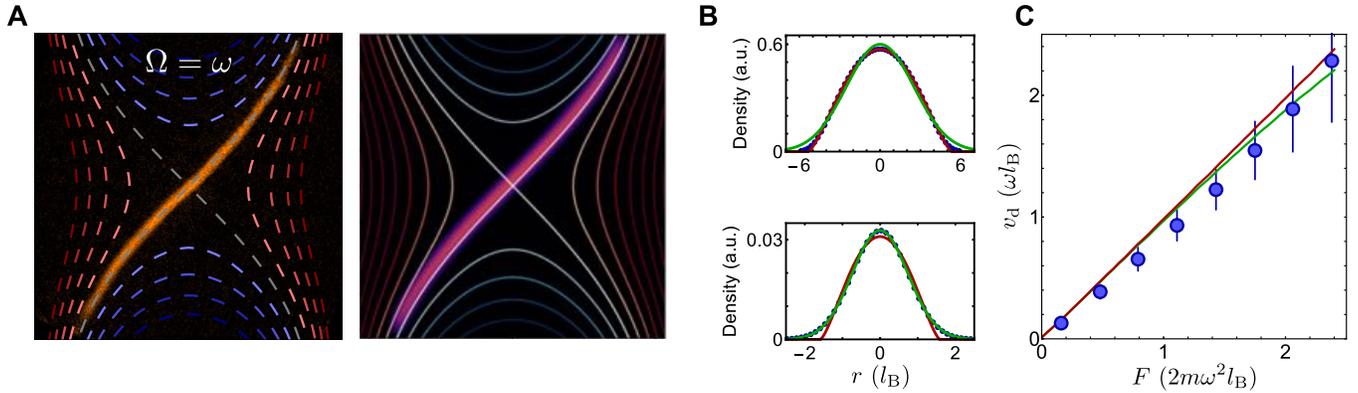

**Fig. S3. Gross-Pitaevskii simulation of the geometric squeezing experiment.** **(A)** Quartic corrections to the TOP trap potential [1] lead to a curvature of the cloud as the atoms flow out along isopotentials. We show images from the experiment (left, taken from Fig. 1 of the main paper) and the simulation (right). **(B)** At high densities (top), a cut through the density distribution along the short axis of the cloud is better fitted by a Thomas-Fermi function (red) than a Gaussian (green). At low densities (bottom), the gas occupies the LLL and the density is better fitted by a Gaussian. Corresponding experimental profiles are shown in Fig. 3C of the main paper. **(C)** The drift velocity as a function of the azimuthal force $F(r) = m\varepsilon\omega^2 r$ felt by atoms flowing out along the diagonal of the saddle potential. The red line shows the simulated drift velocity for a perfectly harmonic trap, and the green line shows the GP result including corrections to the velocity directed along $x = y$, which arise from quartic terms in the trapping potential. The blue points show our experimental data, plotted in Fig. 2 of the main paper.

## THEORETICAL DESCRIPTION

### Derivation of the Foucault Hamiltonian

The Hamiltonian of a two-dimensional harmonic oscillator viewed in a frame rotating about the $z$-axis is

$$H = \frac{p_x^2 + p_y^2}{2m} + \frac{1}{2} m\omega^2(x^2 + y^2) - \Omega L_z, \tag{2}$$

where $\omega$ is the natural frequency of the pendulum, $m$ is the particle's mass, $p_{x,y}$ are canonical momenta along $x$ and $y$, $L_z = xp_y - yp_x$ is the axial angular momentum, and $\Omega$ is the rotational angular frequency of the reference frame. Although we are describing the quantum problem, we omit hats over all operators for brevity.

The equivalence of Eq. (2) to the Hamiltonian of a charged particle in a magnetic field is easily seen by completing the square for momentum variables. For the case $\Omega = \omega$, where the centrifugal force cancels the trapping force, we obtain

$$H = \frac{(p_x + m\Omega y)^2 + (p_y - m\Omega x)^2}{2m}. \tag{3}$$

Rotation of the reference frame therefore induces an effective vector potential $q\vec{A} = m\vec{\Omega} \times \vec{r}$, corresponding to a uniform magnetic field $qB = q\vec{\nabla} \times \vec{A} = 2m\Omega$. Here $q$ is the particle's charge in the equivalent magnetic problem. In that case, particles perform cyclotron motion at the cyclotron frequency $qB/m = 2\omega$ and with a typical extent set by the magnetic length $\ell_B = \sqrt{\hbar/(qB)}$ [10].

The Hamiltonian of the rotating pendulum given in Eq. (2) is conveniently diagonalized by introducing ladder operators corresponding to the counter-rotating 'cyclotron' mode, and the co-rotating 'guiding center' mode,

$$a = \frac{a_x + ia_y}{\sqrt{2}}, \qquad b = \frac{a_x - ia_y}{\sqrt{2}} \tag{4}$$

where $a_x = \sqrt{\frac{m\omega}{2\hbar}}(x + i\frac{p_x}{m\omega})$ and $a_y = \sqrt{\frac{m\omega}{2\hbar}}(y + i\frac{p_y}{m\omega})$ are ladder operators for the individual $x$- and $y$-oscillators respectively. One finds

$$H = \hbar(\omega + \Omega)a^\dagger a + \hbar(\omega - \Omega)b^\dagger b + \hbar\omega, \tag{5}$$

which describes two decoupled oscillators with natural frequencies $\omega \pm \Omega$. In terms of the cyclotron coordinates $(\xi, \eta)$ and the guiding center coordinates $(X, Y)$ [10],

$$a = \sqrt{\frac{m\omega}{\hbar}}\left(\xi + i\eta\right), \qquad b = \sqrt{\frac{m\omega}{\hbar}}\left(X - iY\right), \tag{6}$$

where $(\xi, \eta)$ and $(X, Y)$ are admixtures of the position and momentum variables of the pendulum bob,

$$\begin{aligned} \xi = \left(\frac{x}{2} - \frac{p_y}{2m\omega}\right), \qquad \eta = \left(\frac{y}{2} + \frac{p_x}{2m\omega}\right) \\ X = \left(\frac{x}{2} + \frac{p_y}{2m\omega}\right), \qquad Y = \left(\frac{y}{2} - \frac{p_x}{2m\omega}\right). \end{aligned} \tag{7}$$

A particle's absolute position $(x, y) = (\xi + X, \eta + Y)$ is given by the sum of a fast, counter-rotating motion in $(\xi, \eta)$-space, and a slow, co-rotating motion in $(X, Y)$-space (see Fig. 1A of the main paper). Since these coordinate pairs each span the phase space of a one-dimensional harmonic oscillator, they do not commute, and

$$[\xi, \eta] = -[X, Y] = i\ell_B^2, \tag{8}$$

where $\ell_B = \sqrt{\hbar/(2m\omega)}$ is the rotational analogue of the magnetic length. In terms of these variables, the pendulum Hamiltonian takes the form

$$H = m\omega(\omega + \Omega)(\xi^2 + \eta^2) + m\omega(\omega - \Omega)(X^2 + Y^2), \tag{9}$$

which reproduces Eq. (1) of the main paper.

## Derivation of the squeezing Hamiltonian

Viewed in a rotating reference frame, the Hamiltonian of an anisotropic harmonic oscillator with trapping frequencies $\omega\sqrt{1 \pm \varepsilon}$ along the $x$- and $y$-directions respectively is

$$H_\varepsilon = \frac{p_x^2 + p_y^2}{2m} + \frac{1}{2}m\omega^2(x^2 + y^2) + \frac{1}{2}m\varepsilon\omega^2(x^2 - y^2) - \Omega L_z, \tag{10}$$

where all quantities are defined as in the previous section. The Hamiltonian is quadratic, and can be decoupled into its normal modes by a gauge transformation, under which wavefunctions transform as $\psi' = G\psi$, where $G = \exp\{-i\kappa xy(m\omega/\hbar)\}$ and $\kappa = \varepsilon\omega/(2\Omega)$. The transformed Hamiltonian is

$$H_\varepsilon' = GH_\varepsilon G^\dagger = \frac{p_x^2 + p_y^2}{2m} + \frac{1}{2}m\omega^2(1 + \kappa^2)(x^2 + y^2) - \Omega(xp_y - yp_x) + \kappa\omega(xp_y + yp_x). \tag{11}$$

We now rescale spatial and momentum variables according to

$$\begin{aligned}
\tilde{x} &= (1 + \kappa^2)^{1/4}\,x, & \tilde{p}_x &= (1 + \kappa^2)^{-1/4}\,p_x \\
\tilde{y} &= (1 + \kappa^2)^{1/4}\,y, & \tilde{p}_y &= (1 + \kappa^2)^{-1/4}\,p_y
\end{aligned} \tag{12}$$

which yields

$$H_\varepsilon' = \sqrt{1 + \kappa^2}\left[\frac{\tilde{p}_x^2 + \tilde{p}_y^2}{2m} + \frac{1}{2}m\omega^2(\tilde{x}^2 + \tilde{y}^2)\right] - \Omega(\tilde{x}\tilde{p}_y - \tilde{y}\tilde{p}_x) + \kappa\omega(\tilde{x}\tilde{p}_y + \tilde{y}\tilde{p}_x). \tag{13}$$

The first two terms are simply the Hamiltonian of a rotating isotropic oscillator (see Eq. (2)). Analogously to the previous section, we introduce scaled ladder operators,

$$\tilde{a} = \frac{\tilde{a}_x + i\tilde{a}_y}{\sqrt{2}}, \qquad \tilde{b} = \frac{\tilde{a}_x - i\tilde{a}_y}{\sqrt{2}} \tag{14}$$

in terms of which the first term of Eq. (13) yields $\hbar\omega\sqrt{1 + \kappa^2}\,(\tilde{a}^\dagger\tilde{a} + \tilde{b}^\dagger\tilde{b} + 1)$, the second term $\hbar\Omega\,(\tilde{a}^\dagger\tilde{a} - \tilde{b}^\dagger\tilde{b})$ and the third term $\hbar\kappa\omega\,(\tilde{a}^\dagger\tilde{a}^\dagger + \tilde{a}\tilde{a} + \tilde{b}^\dagger\tilde{b}^\dagger + \tilde{b}\tilde{b})/2$. We finally obtain

$$\begin{aligned}
H_\varepsilon' = {} & \hbar\left[\lambda_+(\tilde{a}^\dagger\tilde{a} + 1/2) - \frac{\kappa\omega}{2}(\tilde{a}^\dagger\tilde{a}^\dagger + \tilde{a}\tilde{a})\right] \\
& + \hbar\left[\lambda_-(\tilde{b}^\dagger\tilde{b} + 1/2) + \frac{\kappa\omega}{2}(\tilde{b}^\dagger\tilde{b}^\dagger + \tilde{b}\tilde{b})\right].
\end{aligned} \tag{15}$$

The Hamiltonian therefore separates into two decoupled oscillators, each including a squeezing interaction, with natural frequencies $\lambda_\pm = \omega\sqrt{1 + \kappa^2} \pm \Omega$ and squeezing rate $\zeta = \kappa\omega$.

In our case, $\Omega = \omega$ and so $\zeta = \varepsilon\omega/2$ and $\kappa = 0.06$, corresponding to an anisotropy-induced shift in the normal mode frequencies by $\sim 10^{-3}\,\omega$. To an excellent approximation, because $\kappa\omega \ll \lambda_+$ we can neglect squeezing of the cyclotron motion and set $\lambda_+ \to 2\omega$. On the other hand, because $\kappa\omega \gg \lambda_-$ the guiding center motion is completely squeezed, and we set $\lambda_- \to 0$. We also neglect the small ($\sim 0.1\%$) rescaling of spatial and momentum variables. This yields the squeezing Hamiltonian

$$H_s \approx 2\hbar\omega(a^\dagger a + 1/2) + \frac{\hbar\zeta}{2}(b^\dagger b^\dagger + bb), \tag{16}$$

which recovers Eq. (3) of the main paper.

## Squeezing as effecting a gauge transformation

The rotational origin of the gauge field in Eq. (3) gives rise to a Hamiltonian expressed in the symmetric gauge. However, one can equally well express the rotating gas Hamiltonian in the Landau gauge. This gauge transformation is accomplished by defining Landau gauge wavefunctions $\psi_L$, which are related to their symmetric gauge counterparts $\psi_s$ by $\psi_L = U\psi_s$ where $U = e^{im\Omega xy/\hbar}$, and whose time evolution is governed by the transformed Hamiltonian $UHU^\dagger$. In the case $\Omega = \omega$, the symmetric gauge Hamiltonian of Eq. (3) transforms to

$$H = \frac{p_x^2}{2m} + \frac{1}{2}m(2\omega)^2(x - k_y\ell_B^2)^2, \tag{17}$$

where we have assumed wavefunctions of the form $\exp(ik_y y)\psi_{\mathrm{L}}(x)$ due to the translational invariance of Eq. (17).

This Hamiltonian corresponds to a flat dispersion associated with momentum along the $y$-direction, and an effective harmonic oscillator along $x$. Physically, the effective trapping arises from the kinetic energy cost imposed by irrotationality of the condensate in the lab frame, which implies a flow profile in the rotating frame $\vec{v} = (0, -2\omega x)$ [9]. Within the Landau gauge, eigenstates are translationally-invariant along $y$, and in the LLL are Gaussian along $x$ with the density showing a $e^{-1/2}$-radius of $\ell_B/\sqrt{2}$.

Geometric squeezing smoothly evolves the wavefunction from one that is most easily described in the symmetric gauge, to one that is best described in the Landau gauge. As a simple illustration, we calculate the evolution of the non-interacting ground state of the 2D harmonic oscillator under geometric squeezing. We will see how this state coherently evolves into the ground state in the Landau gauge, acquiring a width given by the cyclotron zero-point motion.

The ground state starts out in $|0, 0\rangle$, where $|m, n\rangle$ denotes a state of $m$ quanta of cyclotron oscillation and $n$ quanta of guiding center motion. We treat the case, relevant here, of critical rotation $\Omega = \omega$. Standard tools of quantum optics [11] allow us to write the time-evolved state as

$$|\Psi(t)\rangle = \exp\left(-i\frac{\zeta t}{2}\left(b^\dagger b^\dagger + bb\right)\right)|0, 0\rangle = \frac{1}{\sqrt{\cosh(\zeta t)}} \exp\left(-\frac{i}{2}\tanh(\zeta t)\, b^\dagger b^\dagger\right)|0, 0\rangle .$$

The wavefunction at position $(x, y)$ is $\langle x, y|\Psi(t)\rangle$. Noting that $b^\dagger = z - a$, with $z = x + iy$ written in units of the oscillator length $\sqrt{\hbar/m\omega} = \sqrt{2}\, l_B$, $[z, a] = 0$, and $\langle x, y|0, 0\rangle = \frac{1}{\sqrt{\pi}}\exp(-\frac{1}{2}(x^2 + y^2))$, one finds

$$\langle x, y|\Psi(t)\rangle = \frac{1}{\sqrt{\pi \cosh(\zeta t)}} \exp\left(-\frac{i}{2}\tanh(\zeta t)z^2\right) \exp\left(-\frac{1}{2}|z|^2\right), \tag{18}$$

from which follows for the density $n(x, y) = |\langle x, y|\Psi(t)\rangle|^2$

$$n(x, y) = \frac{1}{\pi \cosh \zeta t} \exp\left(-\frac{1}{2}\left(1 - \tanh(\zeta t)\right)(x + y)^2 + \frac{1}{2}\left(1 + \tanh(\zeta t)\right)(x - y)^2\right). \tag{19}$$

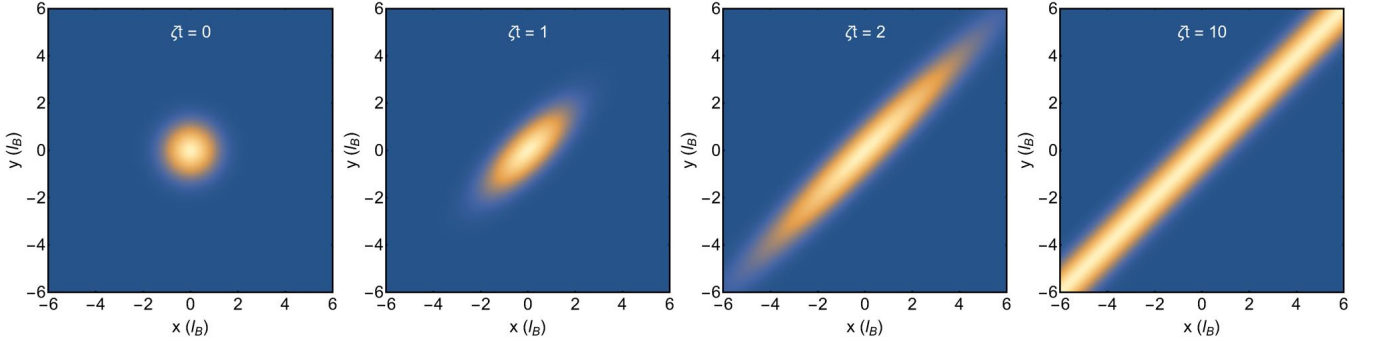

**Fig. S4. Geometric squeezing from a symmetric to Landau gauge ground state.** The density plots show the time evolution of the condensate density $n(x, y)$ according to Eq. (19). The cloud is coherently transformed from the circular ground state wavefunction of the symmetric gauge, to the elongated Landau gauge ground state.

This describes geometric squeezing along the diagonal $x = y$, and the evolution of $n(x, y)$ is illustrated in Fig. S4. The final state for $\zeta t \gg 1$ evolves towards

$$|\Psi(\zeta t \gg 1)\rangle \propto \exp\left(-\frac{1}{2}(x - y)^2\right) \exp\left(-\frac{1}{2}i(x^2 - y^2)\right) \tag{20}$$

which is the ground state Landau gauge wavefunction oriented along the diagonal $y = x$, of density $n(x, y) \propto \exp(-(x - y)^2)$, featuring a $e^{-1/2}$-radius along the narrow axis of $l_B/\sqrt{2}$. The phase profile $\exp(-\frac{1}{2}i(x^2 - y^2)) = \exp(-im\omega uv/\hbar)$, with $u = (x + y)/\sqrt{2}$ and $v = (x - y)/\sqrt{2}$, is precisely that of a Landau gauge strip when expressed in the symmetric gauge (in which we solved the problem) for $\Omega = \omega$.

## Deviations from ideal behavior

The motion of a particle in a rotating frame deviates from the ideal picture of circular cyclotron orbits drifting along isopotentials in two ways [12]. First, any applied force not only leads to a transverse drift in $(X, Y)$-space, but also a static colinear displacement in $(\xi, \eta)$-space. Second, curvature of an applied scalar potential mixes the cyclotron and guiding center modes, resulting in sheared cyclotron orbits.

In our case, atoms are driven outward along the $x = y$ diagonal of the saddle potential by the azimuthal force $F(r) = m\varepsilon\omega^2 r$. Cyclotron orbits occur at a frequency $2\omega$ giving an effective spring constant $k = 4m\omega^2$, which implies a spatial displacement in the azimuthal direction of $F/k = \varepsilon r/4$. This causes the outward flow of atoms to deviate from the $x = y$ diagonal by an angle of $\varepsilon/4 \sim 2°$.

The shearing of cyclotron orbits follows from Eq. (15). Rewriting the cyclotron ladder operator in terms of spatial cyclotron variables, $a = \sqrt{m\omega/\hbar}\,(\xi + i\eta)$, yields a cyclotron Hamiltonian $2m\omega^2[\xi^2(1 - \kappa/2) + \eta^2(1 + \kappa/2)]$. Atoms therefore perform elliptical trajectories in cyclotron phase-space, of ellipticity $\kappa/2 = \varepsilon/4 = 0.03$, giving a ratio of major to minor orbit widths of 1.03.

## Relation of the density distribution to cyclotron and guiding center Wigner functions

In this section, we show that if the particle's wavefunction is separable in guiding center and cyclotron coordinates, then the imaged density distribution is given by the convolution of the guiding center and cyclotron Wigner distributions. In the following we set $\hbar = 1$ and $2m\omega = 1$.

For a one-dimensional particle described by state vector $|\psi\rangle$ and with position and momentum variables $(x, p)$, the Wigner distribution is defined by

$$\mathcal{W}(x, p) = \frac{1}{\pi} \int \mathrm{d}u \, e^{2ipu} \, \psi(x - u) \, \psi^*(x + u), \tag{21}$$

and is closely analogous to the phase space distribution of classical physics [13]. Whereas $[x, p] = i$, the cyclotron and guiding center coordinates instead have commutators $[\xi, \eta] = i$ and $[X, Y] = -i$ (see Eq. (8)), implying corresponding Wigner functions

$$\mathcal{W}_c(\xi, \eta) = \frac{1}{\pi} \int \mathrm{d}u \, e^{2i\eta u} \, \psi_c(\xi - u) \, \psi_c^*(\xi + u),$$

$$\mathcal{W}_g(X, Y) = \frac{1}{\pi} \int \mathrm{d}u \, e^{-2iYu} \, \psi_g(X - u) \, \psi_g^*(X + u). \tag{22}$$

where $\psi_c(\xi)$ and $\psi_g(X)$ are the cyclotron and guiding center wavefunctions.

The imaged density is proportional to the particle state vector $|\Psi\rangle$ projected onto a state of definite $x$ and $y$,

$$
\begin{aligned}
n(x, y) &= |\langle x, y|\Psi\rangle|^2 \\
&= \left| \int\int \frac{\mathrm{d}p_y}{2\pi} \, \langle x, y|x, p_y\rangle \, \Psi(x, p_y) \right|^2 \\
&= \left| \int\int \frac{\mathrm{d}p_y}{2\pi} \, e^{ip_y y} \, \Psi\left(X = \frac{x}{2} + p_y, \, \xi = \frac{x}{2} - p_y\right) \right|^2,
\end{aligned}
\tag{23}
$$

where we have inserted the resolution of the identity $I = \int \frac{\mathrm{d}x\,\mathrm{d}p_y}{2\pi} \, |x, p_y\rangle\, \langle x, p_y|$, and used Eq. (7) to relate absolute spatial and momentum coordinates to cyclotron and guiding center variables.

We now make the assumption of a separable wavefunction, such that $|\Psi\rangle = |\psi_g\rangle\, |\psi_c\rangle$, where $|\psi_g\rangle$ and $|\psi_c\rangle$ are the guiding center and cyclotron state vectors respectively. We find

$$n(x, y) = \frac{1}{4\pi^2} \int\int \mathrm{d}u\,\mathrm{d}v\, e^{i(u-v)y} \, \psi_g\left(\frac{x}{2} + u\right) \psi_g^*\left(\frac{x}{2} + v\right) \psi_c\left(\frac{x}{2} - u\right) \psi_c^*\left(\frac{x}{2} - v\right). \tag{24}$$

Making the coordinate rotation $u = \alpha + \beta, \, v = \alpha - \beta$ yields

$$n(x, y) = \frac{1}{2\pi^2} \int\int \mathrm{d}\alpha\,\mathrm{d}\beta\, e^{2i\beta y} \, \psi_g\left(\frac{x}{2} + \alpha + \beta\right) \psi_g^*\left(\frac{x}{2} + \alpha - \beta\right) \psi_c\left(\frac{x}{2} - \alpha - \beta\right) \psi_c^*\left(\frac{x}{2} - \alpha + \beta\right). \tag{25}$$

We now employ the convolution theorem $\int \frac{\mathrm{d}k}{2\pi} e^{ikx} f(k)g(k) = [\int \frac{\mathrm{d}k}{2\pi} e^{ikx} f(k)] * [\int \frac{\mathrm{d}k}{2\pi} e^{ikx} g(k)]$, where the convolution is defined by $f(x) * g(x) = \int \mathrm{d}u\, f(\frac{x}{2} + u)\, g(\frac{x}{2} - u)$, yielding

$$n(x,y) = \int \frac{\mathrm{d}\alpha}{2\pi} \left[ \frac{1}{\pi} \int \mathrm{d}\beta\, e^{2i\beta y}\, \psi_g \left( \frac{x}{2} + \alpha + \beta \right)\, \psi_g^* \left( \frac{x}{2} + \alpha - \beta \right) \right] *_y$$

$$\left[ \frac{1}{\pi} \int \mathrm{d}\beta\, e^{2i\beta y}\, \psi_c \left( \frac{x}{2} - \alpha - \beta \right)\, \psi_c^* \left( \frac{x}{2} - \alpha + \beta \right) \right]$$

$$= \int \frac{\mathrm{d}\alpha}{2\pi} \mathcal{W}_g(\frac{x}{2} + \alpha, y) *_y \mathcal{W}_c(\frac{x}{2} - \alpha, y) \tag{26}$$

$$= \mathcal{W}_g(x,y) *_{x,y} \mathcal{W}_c(x,y), \tag{27}$$

where $*_i$ denotes convolution over the $i$-direction.

In the case that the cloud occupies the lowest Landau level, the cyclotron wavefunction corresponds to the ground state of a harmonic oscillator with coordinates $(\xi, \eta)$ and natural frequency $2\omega$, and $\langle \xi | \psi_c \rangle \sim \exp(-\xi^2/(2\ell_B^2))$. Evaluating the Wigner function according to Eq. (22) yields $\mathcal{W}_c(\xi, \eta) \sim \exp(-(\xi^2 + \eta^2)/\ell_B^2)$. The experimentally observed density is therefore the convolution of the guiding center Wigner function with a Gaussian, directly visualizing the Husimi-Q representation of the guiding center wavefunction.

———————



\end{document}